\documentclass[11pt,english]{article}
\usepackage[latin9]{inputenc}%

\usepackage{mathpazo}

\usepackage[letterpaper]{geometry}
\usepackage{xcolor,graphicx}
\usepackage[american]{babel}
\usepackage{array}
\usepackage{float}
\usepackage{multirow}
\usepackage{amsmath,amssymb,amsthm,mathrsfs}
\usepackage{setspace}
\usepackage{esint}
\usepackage{pdfsync} 
\usepackage{booktabs}
\usepackage[compact]{titlesec}
\usepackage[authoryear]{natbib}
\PassOptionsToPackage{normalem}{ulem}
\usepackage{ulem}
\usepackage[unicode=true,
bookmarks=true, bookmarksnumbered=false,bookmarksopen=false,
breaklinks=false,pdfborder={0 0 1},backref=false,colorlinks=false]
{hyperref}
\usepackage{breakurl}
\usepackage{asa_manu}
\bibpunct{(}{)}{;}{a}{}{,}
\usepackage{color}

\usepackage{hyperref}
\hypersetup{colorlinks,citecolor=blue,linkcolor=blue,urlcolor=blue}

\date{}

\begin{document}

\title{\Large Optimal Allocation of Gold Standard Testing under Constrained Availability: Application to Assessment of HIV Treatment Failure}

\author{Tao Liu$^1$, Joseph W Hogan$^1$, Lisa Wang$^1$, Shangxuang Zhang$^1$, Rami Kantor$^2$\\ ~ \\
$^1$Brown University School of Public Health \\$^2$Alpert Medical School of Brown University}
\maketitle

\begin{abstract}

	The World Health Organization (WHO) guidelines for monitoring the effectiveness of HIV treatment in resource-limited settings (RLS) are mostly based on clinical and immunological markers (e.g., CD4 cell counts).  Recent research indicates that the guidelines are inadequate and can result in high error rates.  Viral load (VL) is considered the ``gold standard'', yet its widespread use is limited by cost and infrastructure.  In this paper, we propose a diagnostic algorithm that uses information from routinely-collected clinical and immunological markers to guide a selective use of VL testing for diagnosing HIV treatment failure, under the assumption that VL testing is available only at a certain portion of patient visits.  Our algorithm identifies the patient subpopulation, such that the use of limited VL testing on them minimizes a pre-defined risk (e.g., misdiagnosis error rate).  Diagnostic properties of our proposal algorithm are assessed by simulations.  For illustration, data from the Miriam Hospital Immunology Clinic (RI, USA) are analyzed.
\end{abstract}

\vspace*{.3in}
\noindent\textsc{Key  words}: Antiretroviral  failure, constrained optimization, HIV/AIDS, resource limited, ROC, tripartite classification.

\vfill{}

\newpage
\setcounter{page}{1}

\mbox{}
\vspace*{.5in}
\begin{flushleft}
  \textbf{Authors:}
\end{flushleft}

Tao Liu, Ph.D. 

Joseph W.\ Hogan, Sc.D. 

Lisa Wang, Sc.M.  

Shangxuan Zhang, M.S.

Rami Kantor, M.D. 

\mbox{}
\vspace*{.1in}
\begin{flushleft}
  \textbf{Author's Footnote:}
\end{flushleft}

\noindent Tao Liu is Associate Professor (E-mail: tliu@stat.brown.edu), Joseph W.\ Hogan is Professor, and Lisa Wang is Graduate Student, Department of Biostatistics, Center for Statistical Sciences, Brown University School of Public Health, Providence, RI 02912.  Shangxuan Zhang is Statistical Programmer, Memorial Sloan-Kettering Cancer Center, New York City, NY 10016.  Rami Kantor is Associate Professor of Medicine, Division of Infectious Diseases, the Alpert Medical School of Brown University, Providence, RI 02912.  This research is funded by a 2009 developmental grant from the Lifespan/Tufts/Brown Center for AIDS Research (CFAR).  The project described is supported by Grant Number P30AI042853 from the National Institute of Allergy and Infectious Diseases (NIAID).  Work by Dr.\ Kantor is also supported by a grant (Number R01AI66922) from the National Institute of Health (NIH).  The content is solely the responsibility of the authors and does not necessarily represent the official views of the NIAID or NIH.  The authors are grateful for the helpful comments from reviewers, the associate editor, and the editor.  The authors also thank Ms.\ Allison K.\ DeLong for discussions and comments on early versions of the manuscript.

\newpage

\section{Introduction}
\label{sec:intro}

    According to a recent report of the World Health Organization (WHO) \citep{WorldHealthOrganization2010}, almost 40 million people world-wide are infected with \emph{human immunodeficiency virus} (HIV).  Among them, over 97\% live in resource-limited settings (RLS), particularly in sub-Saharan Africa (UNAIDS \citeyear{UNAIDS2010}).  Although the number of people living with HIV remains high, the mortality rate due to acquired immune deficiency syndrome (AIDS) has started to decline since 2006 \citep{UNAIDS2009}, due in large part to the successful rollout of HIV antiretroviral treatment (ART) in RLS \citep{WHO2010}. 
    
    With more and more people having access to ART, treatment failure is inevitable and must be anticipated.  Treatment failure occurs when antiretroviral medications fail to control HIV replication in infected patients.  Common causes of treatment failure include lack of proper medication adherence and development of drug resistance.  The former may be addressed by reinforcing adherence \citep{Gardner2009}, while the latter usually mandates a switch to a more effective ``next line'' ART regimen (e.g., from a first- to a second-line regimen).
    
    Monitoring the effectiveness of HIV treatment and correctly diagnosing treatment failure in a timely manner is critical for preventing HIV-related morbidity and mortality and transmission of the virus.  Incorrect diagnosis of treatment failure can lead to undesired consequences and compromise the success that has been achieved by rolling out ART in RLS.  Specifically, failure to diagnose treatment failure can result in continued viral replication, deterioration of patient's immune system, extra clinical costs such as treatment of opportunistic infections, increased risk of HIV transmission, selection of resistant strains, and death \citep{Anderson2006, Calmy2007, Vekemans2007}.  Meanwhile, incorrectly diagnosing patients as having treatment failure when in fact they do not can prompt a premature switch to the next-line ART.  This generates unnecessary financial burden (second-line therapies cost up to ten times more than first-lines) and potentially accelerates progression toward resistance to next-line therapies, which are most probably the last line in RLS \citep{Vekemans2007}.
    
    In resource-rich countries such as those in much of western Europe and North America, viral load (VL) testing is routine for HIV treatment monitoring (\citealt{Thompson2010}; DHHS \citeyear{DHHS2011}).  In this paper, VL refers to the amount of HIV in the blood as measured using nucleic acid amplification \citep{Hammer2006}. It is a marker that directly reflects the effectiveness of HIV treatment.  Although HIV cannot be eradicated now, patients with adequate adherence can be expected to have \emph{viral suppression}, which generally means that VL is below the lower detection limit of the assay being used (assays used for clinical purposes have lower detection limits of between 20 and 1000 copies/mL).  A patient on adequate ART who has detectable VL after having previously reached an undetectable level is said to have \emph{virological treatment failure} (hereafter ``viral failure'' or ``treatment failure''), an indication that the particular treatment regimen may no longer be effective.
    
    In RLS, VL testing is either limited or not available due to factors such as cost, lack of facilities, and lack of properly trained personnel \citep{Fiscus2006, Calmy2007, Schooley2007}.  Therefore, diagnosis of HIV treatment failure is commonly made using lower-cost and less accurate markers such as current CD4 cell count, CD4 percent among all lymphocytes, and relative changes in these measures since last visit; and clinical indicators such as opportunistic infections, weight loss, and HIV-related malignancies.  Indeed, these immunological and clinical markers form the basis of HIV treatment monitoring guidelines as recommended by the WHO \citep{Calmy2007, WorldHealthOrganization2010}. These guidelines are widely adopted by countries in sub-Saharan Africa (e.g., Malawi \citeyear{Malawi2003a}; Uganda \citeyear{Uganda2003}; Zambia \citeyear{Zambia2004a}; Kenya \citeyear{Kenya2005}) and other developing regions.
    
    Although CD4-based markers are generally associated with VL, a consensus has been reached recently that their use for diagnosing HIV treatment failure is prone to high misclassification rates \citep{Deeks2000, Deeks2002, Moore2005, Bisson2006, Schechter2006, Tuboi2007, Bisson2008, Mee2008, Castelnuovo2009, Kantor2009, Keiser2009,Meya2009,Reynolds2009, Kiragga:2012aa}.  Data from a recent study of patients receiving care through the Academic Model Providing Access to Healthcare (AMPATH) in western Kenya show that almost 40\% of those having treatment failures would have been incorrectly diagnosed based on the WHO guidelines \citep{Kantor2009}.
    
    Several studies have investigated monitoring HIV treatment using markers in addition to or instead of CD4 cell count \citep{Bagchi2007, Kantor2009, Foulkes2010, Abouyannis2011}.  \citet{Bagchi2007} showed that weight loss is associated with treatment failure but pointed out that its clinical utility is limited because weight is influenced by many factors.  \citet{Kantor2009} found in a Kenyan cohort that time on therapy and change in CD4 percent can be potentially incorporated into CD4-based rules to improve the diagnosis of treatment failure.  \citet{Abouyannis2011} developed and tested a scoring system that incorporates CD4 count, mean cell volume, medication adherence, and HIV-associated clinical events for diagnosing treatment failure.  \citet{Foulkes2010} proposed a prediction-based classification method that combines multiple time-varying clinical measures for predicting treatment failure.  Each of these studies focuses on augmenting or replacing CD4 count with other immunological and clinical markers, assuming that VL testing is completely unavailable.  Potential improvements are demonstrated, but often found to be marginal.
    
    In this paper, we consider augmenting rules of diagnosing treatment failure based on low-cost markers (such as CD4 cell count) with a \emph{selective use} of VL testing, under the assumption that VL testing can be ordered only for a fixed portion of patient visits.  Our approach is motivated by the fact that several HIV care programs in developing countries have started to conduct VL testing for some of their patients.  For example, as a result of the study by \citet{Kantor2009}, AMPATH is currently conducting VL testing at about ten percent of its patient visits when treatment failure is suspected.  Our approach is also motivated by the expectation that as technology and training advance~\citep[e.g.,][]{Greengrass2009}, VL testing will be more affordable, even if substantially limited in the near future.
    
    Assuming that VL testing is available but at a fixed portion of patient visits, we propose a tripartite classification procedure to triage VL testing based on a risk score $S$ derived from low-cost non-VL markers.  Specifically, the resulting tripartite diagnostic rule comprises two cut-off values $l$ and $u$ on $S$, with $l\le u$, that classify HIV patients into three mutually exclusive categories (refer to Figure~\ref{fig:Decision-rules-and}), and correspondingly takes one of the following three actions for each category. 
\begin{enumerate}
\item[\textbf{(a)}:] Those with $S>u$ are diagnosed as failing treatment,
\item[\textbf{(b)}:] Those with $S\le l$ are diagnosed as non-failing, and
\item[\textbf{(c)}:] Those with $l<S\le u$ are designated for VL testing, which will provide an error-free diagnosis.
\end{enumerate}

    The tripartite diagnostic rule is designed to minimize a pre-specified risk (e.g., misclassification) subject to the constraint on the availability of VL assays.  To identify the optimal rule, we develop both nonparametric and semiparametric approaches to inference about $l$ and $u$.  We also develop a receiver operating characteristic (ROC) analysis procedure for a general assessment of candidate tripartite rules.  The ROC curve and the area under the ROC curve (AUC) provide a comprehensive measure of diagnosis capacity of tripartite rules, and allow us to evaluate the potential improvement that can be achieved by increasing VL testing availability.  ROC analysis of tripartite rules has many statistical properties that are similar to conventional ROC analysis of bipartite rules.

    The rest of the paper is organized as follows: Notations, definitions, and criteria for rule development are given in Section~\ref{sec:Notations}; nonparametric and semiparametric approaches to optimal rule selection are presented in Section~\ref{sec:Optimal-Rule-Selection}; ROC analysis of tripartite rules is described in Section~\ref{sec:Method-and-Theory}; and simulation studies are conducted in Section~\ref{sec:Simulation-Studies}. For illustration, data from the HIV Immunology Clinic of the Miriam Hospital (RI, USA) are analyzed in Section~\ref{sec:Application}. We conclude with a summary and discussion of future research in Section~\ref{sec:Discussion}.

\section{Notations and Definitions}
\label{sec:Notations}

\subsection{HIV Viral Status and Risk Score}
\label{sub:Risk-Score-and}

    The objective of HIV treatment monitoring is to diagnose viral failure.  Let $V$ denote a patient's (possibly unmeasured) viral load.  Viral failure is said to occur when $V$ exceeds a pre-specified threshold $v^{*}$, where $v^{*}$ is typically the lower detection limit of the VL assay being used.  Let $Z = \mathbf{1} (V>v^{*})$ denote viral status with $Z=1$ indicating a viral failure and $0$ otherwise, where $\mathbf{1}(\cdot)$ is the indicator function.  The prevalence of viral failure is denoted by $p=\Pr(Z=1)$.  At each patient encounter, a set of immunological, clinical, and demographic markers is usually collected, which may include CD4 count, CD4 percent among all lymphocytes (and recent changes in both), WHO stage, time on therapy, hemoglobin, weight, age, gender, and adherence measures.  Henceforth, these markers are generically referred to as low-cost \emph{clinical markers} and denoted by a vector $\mathbf{X}$.

    For each individual, these clinical markers are translated into a scalar risk score $S=S(\mathbf{X})$.  Several recent studies have proposed versions of $S(\mathbf{X})$ for determining the risk of treatment failure \citep[e.g.,][]{Lynen2009, Meya2009,Abouyannis2011}.  If $S(\mathbf{X)}$ is a predicted probability of viral failure given $\mathbf{X}$, it can be derived using logistic regression, regression trees, or other types of prediction-based classification methods \citep[e.g.,][]{Pepe2000, Hastie2001, Foulkes2010, Justice:2010, Van-Der-Laan2011}.  In this paper, we assume that the functional form of $S(\cdot)$ is known, but note that finding and validating an optimal form of $S(\mathbf{X})$ is an important topic of research \citep[see][]{Huang2007, Pepe2008, Steyerberg2010, Pepe2011}

\begin{figure}
  \caption{Risk score distributions and diagnosis actions.}
  \label{fig:Decision-rules-and}
  \begin{centering}
    \includegraphics*[width=.7\textwidth]{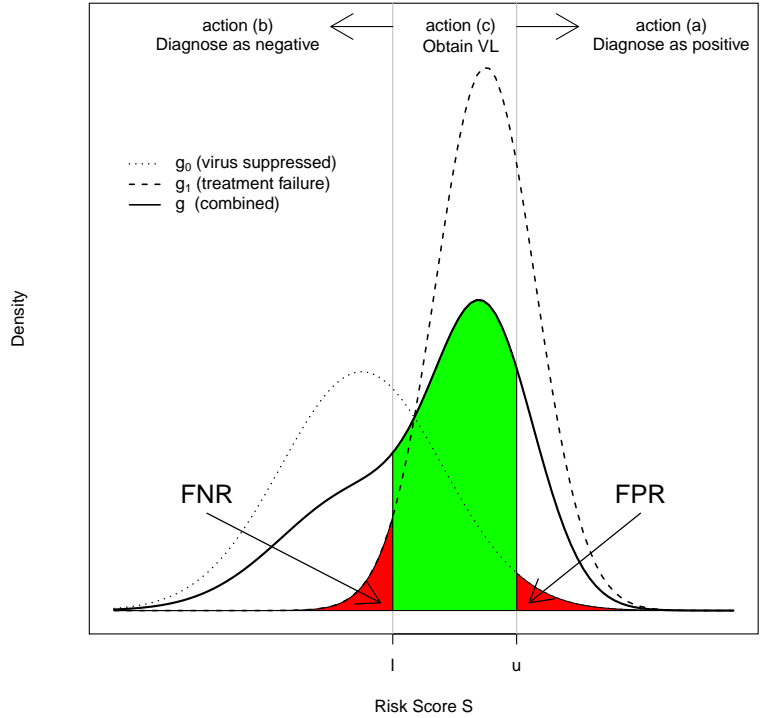}
    \par\end{centering}
\end{figure}

    Let $G_1$ and $ G_0$ denote the distributions of $S$ for patients with viral failure ($Z=1$) and viral suppression ($Z=0$), and $g_{1}$ and $g_{0}$ denote their associated densities, respectively.  The population distribution of $S$ is therefore a mixture distribution $ G=(1-p) G_0+pG_1$, whose density is denoted by $g$.  We assume that for independent observations $S$ and $S'$, where $S\sim G_1$ and $S'\sim G_0$, $S$ is stochastically greater than $S'$ in the sense that on average, patients with viral failure have higher risk scores.  An illustration of $g_{0}$, $g_{1}$, and $g$ leading to a hypothetical distribution of $S$ is presented in Figure~\ref{fig:Decision-rules-and}.

\subsection{Classification Cut-offs and Tripartite Rules}
\label{sub:Tripartite-Classification-Rules}

	The tripartite diagnostic rule can be formalized as follows. Let $l$ and $u$, with $l\le u$, subdivide the population into three categories: those whose risk of treatment failure is high ($S>u$), low ($S\le l$), or intermediate ($S \in \mathscr{I}\equiv(l,u]$).  Let $\delta_{\mathscr{I}}(S)$ denote the diagnostic decision based on $S$, with $\delta_{\mathscr{I}}(S) = 1$ indicating a treatment failure diagnosis and $\delta_{\mathscr{I}}(S) = 0$ a non-failing diagnosis.  Then our tripartite rule is expressed as
\begin{equation} 
  	\label{eq:rule.2cutoff} 
	\delta_{\mathscr{I}}(S)=\left\{ 
	\begin{array}{ll} 0 & \textrm{if \ensuremath{S\le l}},\\ Z & \textrm{if }S\in\mathscr{I},\\ 1 & \textrm{if \ensuremath{S>u}}.  
	\end{array}\right.  
\end{equation} 
This rule obtains the gold standard measurement for the intermediate risk subpopulation $\{S\in\mathscr{I}\}$, which carries the greatest uncertainty about true viral status.  Note that when $S\in \mathscr{I}$, the diagnosis decision corresponds to the true viral failure status and therefore leads to a correct diagnosis.

\subsection{Loss and Risk Functions}
\label{sub:Constraint-under-a}

	Let $L(d,z)$ denote the loss or cost incurred when the true viral failure status is $Z=z$ and a diagnostic decision $d$ is taken.  Two commonly used loss functions in studies of medical diagnosis are $L_{1}(d,z) = \mathbf{1}(d \ne z)$, which indicates whether a misdiagnosis occurs, and $L_{2}(d,z) = \{\mathbf{1}(d=0,z=1), \mathbf{1}(d=1,z=0)\}^{\top}$, which indexes misdiagnoses separately for those with viral failure (i.e., false negative, FN) and those without (i.e., false positive, FP).  Loss functions can be made more elaborate and extended to incorporate potential costs as well as benefits of correct and incorrect diagnoses (e.g., expected mortality, cost of switching to next-line therapies, and gain of life expectancy); see \citet{Parmigiani2002} for further discussions.

	The development of our diagnostic rule also uses a weighted loss function 
\begin{equation*} 
  L_3(d,z;\lambda) = \lambda\mathbf{1}(d=0,z=1) + (1-\lambda) \mathbf{1}(d=1,z=0), 
\end{equation*} 
where $\lambda \in [0,1]$ is a user-specified weight that reflects relative loss for the two types of misdiagnoses.  At the extremes, setting $\lambda=1$ places the highest priority on avoiding FN (incorrectly diagnosing a patient as non-failing), while $\lambda=0$ prioritizes avoidance of FP (incorrectly diagnosing a patient as treatment failure).  An appropriate and meaningful value of $\lambda$ should be contextually specific and take into account the available information about patient's health status and various costs associated with FP and FN.

	The overall diagnostic accuracy of a diagnostic rule is summarized by a risk function defined as $R(\mathscr{I}) = \mathrm{E}[L(\delta_{\mathscr{I}} (S),Z)]$, where the expectation is taken over the joint distribution of $(S,Z)^{\top}$ \citep{Berger:1985}.  For the loss function $L_{1}$, $R_{1}(\mathscr{I}) = \mathrm{E}[\mathbf{1} \{\delta_{\mathscr{I}}(S) \ne Z\}]$ is the total misclassification rate (TMR).  For $L_{2}$, $R_{2}(\mathscr{I})= \{p\mathrm{FNR}, (1-p)\mathrm{FPR}\}^{\top}$, where FNR and FPR are the FN and FP rates, respectively.  For $L_{3}$, we have a weighted sum of FPR and FNR 
\begin{equation}
  \label{eq:weight.risk}
  R_{3}(\mathscr{I}; \lambda) = \lambda p\mathrm{FNR}+(1-\lambda)
  (1-p)\mathrm{FPR}, 
\end{equation}
where the weights depend on both $\lambda$ and the prevalence of viral failure. Risk function $R_{3}(\mathscr{I};\lambda)$ is one form of `net benefit' functions that have been used in decision curve analyses and utility analyses \citep{Vickers2006, S.G.2009}.  As a special case when $\lambda=.5$, minimizing $R_3(\mathscr{I}; .5)$ is equivalent to minimizing $R_1(\mathscr{I})$.

In the next section, we develop methods for obtaining optimal rules under the risk criteria $R_1 (\mathscr{I})$ and $R_3 (\mathscr{I}; \lambda)$.  The optimal rules that minimize $R_{1}(\mathscr{I})$ and $R_{3}(\mathscr{I};\lambda)$ are called the min-TMR rules and min-$\lambda$ rules, respectively.  In Section \ref{sec:Method-and-Theory}, the vector-valued risk function $R_{2}(\mathscr{I})$ is used to develop a ROC analysis procedure for a general assessment of tripartite diagnostic rules.

\section{Optimal Rule Selection: Constrained Optimization}
\label{sec:Optimal-Rule-Selection}

\subsection{Characterization of Constraints on Gold Standard Testing}
\label{sec:constraint}

Suppose that VL tests can be ordered for a fixed portion $\phi$ of     patient visits, where $0 \le \phi \le 1$.  Then the proposed tripartite rules must satisfy the constraint 
\begin{equation} 
	\label{eq:constraint.rule} 
	G(u)- G(l)\le\phi.  
\end{equation} 
In the extreme cases, $\phi=0$ means that no VL testing is available, while $\phi=1$ means that it is available at all patient visits.

Tripartite diagnostic rules that satisfy \eqref{eq:constraint.rule} can be infinitely many, because if $\delta_{\mathscr{I}}(s)$ satisfies \eqref{eq:constraint.rule}, so does $\delta_{\mathscr{I}'}(s)$ for all $\mathscr{I}'\subset\mathscr{I}$.  We therefore restrict attention only to those rules that take maximum advantage of the available VL tests.  All such rules form our decision space. Specifically, the decision space is defined as the set $\mathscr{D}_{\phi}^{ G}=\{\delta_\mathscr{I}: G(u)- G(l)\le\phi \}$ with the condition that for any $\delta_{\mathscr{I}}(s) \in \mathscr{D}_{\phi}^{G}$, there does not exist another rule $\delta_{\mathscr{I}'}(s)$ with $\mathscr{I}' \supset \mathscr{I}$ and satisfying \eqref{eq:constraint.rule}. 

For a given risk function $R(\cdot)$ and a decision space $\mathscr{D}_{\phi}^G$, the optimal rule is defined as
\begin{equation}
  \label{eq:optimal.rule}
  \delta_{\mathscr{I}^{*}}=\mathop{\arg\min}_{\delta_{\mathscr{I}}\in 
    \mathscr{D}_{\phi}^{G}}\{R(\mathscr{I})\},
\end{equation}
where $\mathscr{I}^{*}$ indicates the optimal cut-offs on $S$ for triaging the VL tests.  We assume that the optimal rule is unique.

\subsection{Optimal Rule Selection}
\label{sub:Optimal-Rule-Selection}

In this section, we develop nonparametric and semiparametric approaches to determining the optimal rule from $\mathscr{D}_{\phi}^{G}$.  The nonparametric approach places no distributional assumption on either $G_0$ or $G_1$ and can therefore be broadly applied.  The semiparametric approach assumes that $G_0$ and $G_1$ follow an exponential tilt model, whereby the densities $g_{0}(s)$ and $g_{1}(s)$ differ only by a factor proportional to $\exp(\beta_{1}s)$, where $\beta_{1}$ is an unknown scalar parameter (called the tilting parameter).  In Section \ref{sec:Simulation-Studies}, we use simulations to show that when the exponential tilt assumption holds, the semiparametric approach is generally more efficient in estimating the optimal rule when sample size is large.

\subsubsection{Nonparametric Approach.}

Suppose that we have a training data set of independent pairs $(S_{1}, Z_{1}),\ldots, (S_{n}, Z_{n})$.  We first estimate $G_1$, $ G_0$, and $ G$ empirically via
\begin{align*}
  \widehat{G}_{z}(s) & =\frac{\sum_{i=1}^{n}\mathbf{1}(S_{i}\le s,
    Z_{i}=z)}{\sum_{i}^{n}\mathbf{1}(Z_{i}=z)},\, z=0,1,\\
  \widehat{ G}(s) &
  =\widehat{p}\widehat{G}_{1}(s)+(1-\widehat{p})\widehat{G}_{0}(s)
\end{align*}
with $\widehat{p}=\sum_{i=1}^{n}Z_{i}/n$.  To determine the optimal rule using \eqref{eq:optimal.rule}, we then obtain the empirical decision space $\mathscr{D}_{\phi}^{\widehat G}$ by the following steps:
\begin{enumerate}
\item Write $\widetilde{\mathbf{L}}=(\widetilde{l}_{1}, \widetilde{l}_{2},
  \dots, \widetilde{l}_{n})^{\top} = (S_{(1)}, S_{(2)},
  \dots,S_{(n)})^{\top}$, where $\widetilde{l}_{j}=S_{(j)}$ is the $j$-th
  order statistic of $\mathbf{S}=(S_{1}, \dots, S_{n})^{\top}$.
\item For each $\widetilde{l}_{j}$, calculate $\widetilde{u}_{j} =
  \arg\max_{u\in\mathbf{S}} \{\widehat{G}(u)- \widehat{G}
  (\widetilde{l}_{j}) \le \phi\}$.  Let $\widetilde{\mathbf{U}} =
  (\widetilde{u}_{1}, \widetilde{u}_{2}, \dots,
  \widetilde{u}_{n})^{\top}$.
\item For $\widetilde{u}_{j}$ and $\widetilde{u}_{j'} \in
  \widetilde{\mathbf{U}}$, $j<j'$, if $\widetilde{u}_{j} =
  \widetilde{u}_{j'}$, drop $\widetilde{l}_{j'}$ from
  $\widetilde{\mathbf{L}}$ and $\widetilde{u}_{j'}$ from
  $\widetilde{\mathbf{U}}$.  Denote the resulting vectors by
  $\widehat{\mathbf{L}} = (\widehat{l}_{1}, \widehat{l}_{2}, \dots,
  \widehat{l}_{m})^{\top}$ and $\widehat{\mathbf{U}} = (\widehat{u}_{1},
  \widehat{u}_{2}, \dots, \widehat{u}_{m})^{\top}$ with $m\le n$.
\item The empirical decision space is given by
  $\mathscr{D}_{\phi}^{\widehat{G}} = \{\delta_{ \widehat{\mathscr{I}}_{j}} :
  j=1,2, \dots, m\}$ with $\widehat{\mathscr{I}}_{j} = (\widehat{l}_{j},
  \widehat{u}_{j}]$.
\end{enumerate}

With the empirical decision space $\mathscr{D}_{\phi}^{\widehat{G}}$, the optimal rule is then estimated via \eqref{eq:optimal.rule} with $\mathscr{D}_{\phi}^{ G}$ replaced by $\mathscr{D}_{\phi} ^{\widehat{G}}$. This can be carried out using a grid search.  For example, to estimate the optimal rule that minimizes TMR, we calculate $\widehat{\mathrm{FNR}}_j = \widehat{G}_{1}(\widehat{l}_{j})$ and $\widehat{\mathrm{FPR}}_j = 1-\widehat{G}_{0}(\widehat{u}_{j})$ for $j=1, \dots, m$.  Then, the optimal min-TMR rule is the rule in $\mathscr{D}_{\phi}^{\widehat{G}}$ that has a risk equal to $\underset{j}{\min} (\widehat{p}\widehat{\mathrm{FNR}}_j + (1-\widehat{p}) \widehat{\mathrm{FPR}}_j)$.  Similarly, to identify the rule that minimizes $R_{3}(\mathscr{I};\lambda)$ for a pre-specified $\lambda$, we select the rule in $\mathscr{D}_{\phi}^{\widehat{ G}}$ that has a risk of $\underset{j}{\min} (\lambda\widehat{p} \widehat{\mathrm{FNR}}_j + (1-\lambda)(1-\widehat{p}) \widehat{\mathrm{FPR}}_j)$.

\subsubsection{Semiparametric Approach}
\label{sec:semip-appr}

The exponential tilt model has been used to characterize the relationship between components of a mixture distribution \citep{ANDERSON:1972, Anderson1979, PRENTICE:1979, Efron:1981, Qin1999}.  The model places no parametric assumptions on individual components of the mixture, except assuming that they differ only by a factor of the form
\begin{equation}
  \label{eq:exponential.tilt}
  g_{1}(s)=\exp(\beta_{0}^{*}+\beta_{1}s)g_{0}(s),
\end{equation}
where $\beta_{1}$ is an unknown tilting parameter and $\beta_{0}^{*} = -\log\mathrm{E}_{ G_0}(e^{\beta_{1}S})$ is a normalizing constant.  Although no constraints are placed on $g_0$, many commonly-used parametric distribution families can be represented in the form of \eqref{eq:exponential.tilt}, such as binomial, Poisson, normal with a common variance, and gamma distributions with a common shape parameter.  In our case, the exponential tilt model is equivalent to the logistic model
\begin{equation}
  \label{eq:logistic.model}
  \mathrm{logit\{Pr}(Z=1\mid S=s)\}=\beta_{0}+\beta_{1}s
\end{equation}
with $\text{logit}(y) = \log\{y/(1-y)\}$ and $\beta_{0} = \beta_{0}^{*}+\mathrm{logit}(p)$.

When the exponential tilt assumption holds, we can estimate $G_0$ and $G_1$ semiparametrically using the results in Appendix \ref{sub:A.-Semi-parametric-estimate}, and then estimate the optimal rule using a grid-search in a similar way to what has been described in the last section.

If our goal is to identify a rule that minimizes TMR, it turns out that we can readily determine this rule without calculating the semiparametric estimates of $G_0$ and $G_1$. To see this, we write $\Gamma(l,u,\tau) = R_{1} (\mathscr{I}) + \tau(G(u)- G(l)-\phi)$, and apply the Lagrange multiplier to solve $\partial\Gamma / \partial(l,u,\tau)^{\top} = 0$.  It is straightforward to verify that the resulting rule must satisfy
\begin{equation*}
  l+u=-\frac{2\beta_{0}}{\beta_{1}},\; G(u)- G(l) \approx \phi.
\end{equation*}
That is, the optimal interval $\mathscr{I}^{*}$ for triaging the limited VL testing is centered at $-\beta_{0} / \beta_{1}$, independent of the VL test availability $\phi$.  The optimal cut-off values therefore can be estimated by $\widehat l=-\widehat{\beta}_{0} / \widehat{\beta}_{1}-\Delta_\phi$ and $\widehat u=-\widehat{\beta}_{0} / \widehat{\beta}_{1}+\Delta_\phi$, where $\widehat{\beta}_{0}$ and $\widehat{\beta}_{1}$ are parameter estimates of the logistic model \eqref{eq:logistic.model} and
\begin{equation*}
  \Delta_\phi = \mathop{\arg\max}_{s}\{\widehat{G}(-\widehat{\beta}_{0} / 
  \widehat{\beta}_{1}+s) - \widehat{G}(-\widehat{\beta}_{0} / 
  \widehat{\beta}_{1}-s) \le \phi\}.
\end{equation*}
In the above equation, the empirical estimate $\widehat{G}$ is used because the semiparametric estimate of $G$ under the exponential tilt model is the same as $\widehat{G}$.

\subsubsection{Uniqueness of Estimated Optimal Rule.}

The estimated optimal rule based on a finite data set may not be unique, even though the true optimal rule is unique.  When there are multiple rules that meet the optimality criterion, we propose to impose additional secondary criteria so as to determine a single optimal rule.  For example, to determine an optimal rule from multiple rules that equally minimize $R_3(\mathscr{I};\lambda)$, we may consider adding $R_1(\mathscr{I})$ as a secondary criterion and choosing one from these rules that has the lowest TMR.  It is also reasonable to randomly choose one for practical use if the estimated optimal rules differ little.

\section{ROC Analysis}
\label{sec:Method-and-Theory}

ROC analyses have been widely used to assess the overall diagnostic accuracy of bipartite classification rules.  An ROC curve is a graphical presentation of the risk function $R_{2}(\cdot)$ associated with all candidate rules in a decision space.  Comprehensive reviews of ROC analyses in biomedical research can be found in \citet{Pepe2000a}, \citet[Ch 2]{Zhou2002}, \citet[Chs 4-5]{Pepe2003}, and \citet{Gatsonis:2009}.  Recent applications of ROC analyses in studies of HIV-infected populations include \citet{Pahwa2008}, \citet{Joska2011}, and \citet{Mabeya:2012}, among many others.  

\subsection{ROC Curve for Tripartite Rules and AUC}
\label{sub:Receiver-Operation-Characteristics}

ROC analyses for tripartite rules can be carried out in a fashion similar to conventional ROC analyses.  With each rule in $\mathscr{D}_{\phi}^{G}$ represented by a point in a 2-dimensional space with its $(\text{FPR}, \text{1\ensuremath{-}FNR})$ as the coordinates, an ROC curve for tripartite rules can be generated by connecting these points using a non-decreasing curve.  Mathematically, we can express the ROC curve for tripartite rules as
\begin{equation}
  \label{eq:ROC.curve}
  C_{\phi}(t):t\in[0,1] \mapsto 1-G_1\circ H_{\phi} \circ 
   G_0^{-1}(1-t),
\end{equation}
where $G_0^{-1}(t) = \inf\{s:G_0(s) \ge t\}$ is the generalized inverse of a cadlag function, $H_{\phi}(u) = \arg\inf_{w}\{G(u)-G(w) \le \phi\}$, and `$\circ$' denotes the function composition operator.  Note that the difference between \eqref{eq:ROC.curve} and a conventional ROC curve is the operation induced by $H_{\phi}$, which dictates that for each $u$ and resulting FPR, the corresponding FNR is calculated based on a lower cut-off $l = H_{\phi}(u) \le u$.

The area under the ROC curve (AUC) for tripartite rules is defined as,
\begin{equation}
  \label{eq:auc}
  \mathrm{AUC}_{\phi}=\int_{0}^{1}C_{\phi}(t)\mathrm{d}t. 
\end{equation}
Like AUC for bipartite classification rules, $\mathrm{AUC}_\phi$ provides an omnibus measure of diagnostic capability of all candidate rules in $\mathscr{D}_{\phi}^{G}$.  It can be interpreted as the true positive rate averaged across all FNRs.  In Appendix \ref{sub:Properties-of-AUC}, we present several properties of AUC$_{\phi}$, which turn out to be very similar to the AUC from a conventional ROC analysis.

As a special case when $\phi=0$, $C_\phi(t)$ and $\mathrm{AUC}_\phi$ reduce to conventional ROC curve and AUC for bipartite rules.

\subsection{Estimation}

With a training data set of independent $(S_{1},Z_{1}), \ldots, (S_{n},Z_{n})$ and a given $\phi$, we can estimate the ROC curve for tripartite rules nonparametrically by
\begin{equation}
  \label{eq:ROC.curve.est}
  \widehat{C}_{\phi}(t) = 1-\widehat{G}_{1}\circ\widehat{H}_{\phi}
  \circ\widehat{G}_{0}^{-1}(1-t),
\end{equation}
where $\widehat{G}_{0}$ and $\widehat{G}_{1}$ are empirical estimates, and $\widehat{H}_{\phi}(u) = \arg\min_{w} \{\widehat{G}(u) - \widehat{G}(w) \le \phi\}$.  The estimated ROC curve is a step function with jumps only at points representing the rules in $\mathscr{D}_{\phi}^{\widehat{G}}$.  When the exponential tilt assumption holds, the ROC curve also can be estimated semiparametrically by replacing $\widehat{G}_{0}$ and $\widehat{G}_{1}$ in \eqref{eq:ROC.curve.est} by their corresponding semiparametric estimates given in Appendix~\ref{sub:A.-Semi-parametric-estimate}.

Using the results in the Appendix \ref{sub:Properties-of-AUC} (See Eq.\eqref{eq:AUC.result.1a}), we can estimate AUC$_{\phi}$ nonparametrically by
\begin{equation}
  \label{eq:auc.nonpar.est}
  \widehat{\mathrm{AUC}}_{\phi}=\frac{1}{n^{2}\widehat{p}(1-\widehat{p})}
  \sum_{j=1}^{n}\sum_{i=1}^{n}Z_{i}(1-Z_{j})\left[\mathbf{1}\{S_{i}>
    \widehat{H}_{\phi}(S_{j})\}+\frac{\mathbf{1}\{S_{i} = 
      \widehat{H}_{\phi}(S_{j})\}}{2}\right]. 
\end{equation}
In Appendix \ref{sub:Asymptotic-properties}, we present several large-sample properties of the nonparametric estimates $\widehat C_\phi(t)$ and $\widehat{\mathrm{AUC}}_\phi$.

\subsection{Using ROC curve for Rule Selection}

An ROC curve for tripartite rules also can be used for optimal rule selection, recognizing that the diagnostic properties of each rule in $\mathscr{D}_{\phi}^{ G}$ are characterized by a point on the curve.  For example, if the ROC curve is smooth and concave, it can be verified that the min-$\lambda$ rule corresponds to the point on the ROC curve where the tangent is equal to $(1-\lambda)(1-p)/(\lambda p)$ \citep{Metz1978}.  Broader discussions on using ROC curves for optimal rule selection can be found in \citet{Zhou2002} and \citet{Pepe2003}.

\section{Simulation Studies} 
\label{sec:Simulation-Studies}

In this section, we conduct simulation studies to examine 1) the diagnostic accuracy of the optimal rules estimated by the nonparametric and semiparametric approaches, and 2) the large-sample properties of the estimated optimal rules.  For the first aim, we consider two scenarios when the exponential tilt assumption is and is not satisfied.  For the second aim, we focus on the setting where the exponential tilt assumption holds.  For simplicity, we consider estimating the optimal rules that minimize TMR.

We use the negative value of CD4 count as a risk score. We first simulate viral status $Z$ assuming that $Z\sim\mathrm{Bernoulli}(p)$, and then conditional on $Z$, simulate $(\mathrm{CD4}|Z=z)=\lceil W\rceil$ with $W \sim \mathrm{Gamma}(\eta_{z},\kappa_{z})$, where $\lceil \cdot \rceil$ denotes the ceiling operation, and $\eta_{z}$ and $\kappa_{z}$ are the shape and scale parameters of the gamma distribution.

Scenario (A) considers the case when the exponential tilt assumption does not hold. We conduct two simulations by simulating CD4 count data from gamma distributions with parameters,
\begin{eqnarray*}
  & \textrm{(A-1): } & (\eta_{0},\kappa_{0})=(3.2,152)
  \textrm{ and }(\eta_{1},\kappa_{1})=(2.3,133)\\
  & \textrm{(A-2): } & (\eta_{0},\kappa_{0})=(4.8,100)
  \textrm{ and }(\eta_{1},\kappa_{1})=(2.3,133).
\end{eqnarray*}
The parameter values of (A-1) are chosen as the maximum likelihood estimates (MLEs) obtained by fitting gamma distributions to the Miriam Immunology Clinic data (which will be analyzed in Section \ref{sec:Application}).  For (A-2), we choose the same values of $(\eta_{1}, \kappa_{1})$ as in (A-1), but set $(\eta_{0},\kappa_{0})$ such that the exponential tilt assumption is further violated while keeping $\eta_0\kappa_0$ unchanged, i.e.\ the average CD4 count for patients without treatment failure stays the same as (A-1). (Recall that the mean of gamma distribution is $\eta\kappa$.)

\begin{figure}
  \caption{Gamma distributions used for simulating CD4 count data. The gray step lines in the top-left subplot are histograms of the CD4 data from the Miriam Hospital Immunology Clinic.  The smooth dashed (solid) lines are gamma densities for those with (without) treatment failure.}
  \label{fig:Gamma-densities}
  \begin{centering}
    \includegraphics[width=0.75\textwidth]{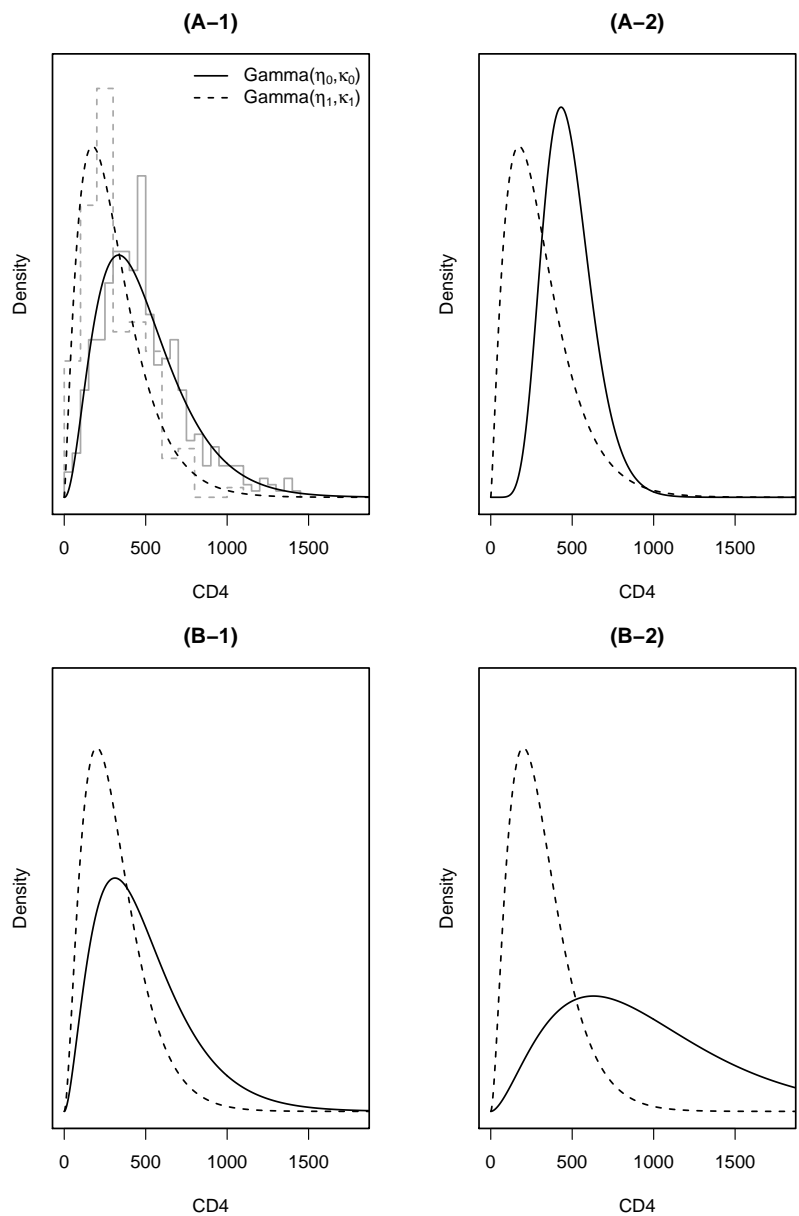}
    \par\end{centering}
\end{figure}

Scenario (B) considers the case when the exponential tilt assumption holds.  We choose a common shape parameter $\eta_{0} = \eta_{1} = 2.8$, the mid-point of $\eta_{0}$ and $\eta_{1}$ in (A-1), and conduct two simulations with parameters 
\begin{eqnarray*} 
& \textrm{(B-1): } & (\eta_{0},\kappa_{0})=(2.8,173) \textrm{ and }(\eta_{1},\kappa_{1})=(2.8,111)\\ 
& \textrm{(B-2): } & (\eta_{0},\kappa_{0})=(2.8,350) \textrm{ and }(\eta_{1},\kappa_{1})=(2.8,111).  
\end{eqnarray*} 
The values of $\kappa_0$ and $\kappa_1$ in (B-1) are the MLEs obtained by fitting gamma distributions to the Miriam Immunology Clinic data while fixing their shape parameters at 2.8.  For (B-2), we choose a large value of $\kappa_{0}=350$ to simulate a case when two gamma distributions are further separated. The gamma densities of the four simulations are shown in Figure \ref{fig:Gamma-densities}.


\subsection{Diagnostic Accuracy}

For the first aim, we consider three prevalences of treatment failure, $p=(.15,.25,.40)$, and assume that VL testing is available at proportions $\phi=(0,20,40)\%$ of patient visits.  For each parameter combination, we simulate 1000 datasets each having 5000 observations.  The first 2500 observations of each dataset are used as the training data to develop an optimal rule, and the remaining 2500 observations as the testing data to calculate its associated misclassification rate.  Results are shown in Table \ref{tab:sim.rules.no.expo.tilt}.

When the prevalence of treatment failure is low (e.g., $p \le .25$) and VL test availability $\phi$ is high, the semiparametric approach may yield a negative estimate of the lower cut-off value on CD4 count, particularly when the center of the optimal cut-off interval is close to zero.  When this occurs, we replace the negative cut-off values by zero.  This adjustment does not imply that the algorithm fails.  It can be verified that the upper cut-off estimate is still correct, and the optimal rule in this case is to assign VL test to those high-risk patients with CD4 count less than the upper cut-off value.  

Table \ref{tab:sim.rules.no.expo.tilt} shows that the nonparametric approach yields the correct estimates of the optimal cut-off values for both Scenarios (A) and (B), and the resulting TMRs are close to the underlying truth.  When the exponential tilt assumption does not hold as in Scenario (A), the optimal rules estimated by the semiparametric approach are slightly biased (contrasted with Scenario (B)).  However when the exponential tilt assumption holds as in Scenario (B), the semiparametric approach yields the correct estimates of the optimal cut-off values, and the estimated cut-off values have much smaller standard errors compared with their corresponding nonparametric estimates.

\begin{table}
  \caption{Simulation results.  For each condition, the estimated lower and upper cut-off values are averages over 1000 simulations, and converted to the original scale of CD4 count.  The numbers in parentheses are standard errors.}
  \label{tab:sim.rules.no.expo.tilt}
  {\footnotesize
  \begin{tabular}{ccccccccccccccc}
    \toprule
    \multirow{2}{*}{} & \multirow{2}{*}{$p$} & \multirow{2}{*}{$\phi$} &  & \multicolumn{3}{c}{true values} &  & \multicolumn{3}{c}{nonparametric estimate} &  & \multicolumn{3}{c}{semiparametric estimate}\\
    \cmidrule{5-7} \cmidrule{9-11} \cmidrule{13-15} 
 &  &  &  & lower & upper & TMR  &  & lower & upper & TMR &  & lower & upper & TMR\\
 \midrule
(A-1) & .15  & 0  &  & 65  & 65  & .15  &  & 70 (20) & 70 (20) & .148  &  & 0 (1) & 0 (1)  & .150 \\
 &  & .2  &  & 0  & 230  & .09  &  & 29 (14) & 231 (5) & .086  &  & 0 (0) & 230 (5) & .086 \\
 &  & .4  &  & 0  & 348  & .05  &  & 25 (13) & 349 (6) & .051  &  & 0 (0) & 348 (6) & .051 \\
 & .25  & 0  &  & 125  & 125  & .24  &  & 128 (23) & 128 (23) & .237  &  & 79 (25) & 79 (25) & .240 \\
 &  & .2  &  & 66  & 225  & .15  &  & 71 (21) & 228 (9) & .152 &  & 2 (8) & 214 (5) & .153 \\
 &  & .4  &  & 39  & 333  & .09  &  & 47 (20) & 336 (7) & .092 &  & 0 (0) & 329 (6) & .092 \\
 & .4  & 0  &  & 239  & 239  & .32  &  & 240 (28) & 240 (28) & .327 &  & 268 (13) & 268 (13) & .327 \\
 &  & .2  &  & 177  & 288  & .23  &  & 192 (22) & 302 (22) & .233 &  & 212 (13) & 323 (14) & .234 \\
 &  & .4  &  & 149  & 378  & .15  &  & 148 (22) & 377 (22) & .153 &  & 153 (14) & 382 (14) & .152 \\
(A-2) & .15  & 0  &  & 130  & 130  & .14  &  & 132 (16) & 132 (16) & .135 &  & 35 (23) & 35 (23) & .147 \\
 &  & .2  &  & 72  & 268  & .07  &  & 71 (20) & 268 (6) & .075 &  & 0 (0) & 262 (5) & .075\\
 &  & .4  &  & 57  & 373  & .04  &  & 59 (21) & 374 (6) & .045 &  & 0 (0) & 370 (5) & .045\\
 & .25  & 0  &  & 176  & 176  & .21  &  & 178 (17) & 178 (17) & .209 &  & 148 (17) & 148 (17) & .212 \\
 &  & .2  &  & 122  & 270  & .13  &  & 121 (17) & 269 (9) & .131 &  & 51 (28) & 247 (7) & .135\\
 &  & .4  &  & 98  & 368  & .08  &  & 96 (18) & 368 (8) & .080 &  & 1 (3) & 351 (5) & .083\\
 & .4  & 0  &  & 249  & 249  & .29  &  & 251 (20) & 251 (20) & .291  &  & 290 (10) & 290 (10) & .295\\
 &  & .2  &  & 200  & 312  & .20  &  & 203 (17) & 314 (15) & .201 &  & 236 (11) & 345 (11) & .204 \\
 &  & .4  &  & 167  & 394  & .13  &  & 169 (18) & 396 (15) & .130  &  & 177 (11) & 403 (10) & .130 \\
\midrule 
(B-1)  & .15  & 0  &  & 0  & 0  & .15  &  & 27 (15) & 27 (15) & .150  &  & 0 (0) & 0 (0) & .150 \\
 &  & .2  &  & 0  & 220  & .10  &  & 20 (9) & 221 (5)  & .095  &  & 0 (0) & 221 (5)  & .095 \\
 &  & .4  &  & 0  & 338  & .05  &  & 18 (8) & 338 (6) & .056  &  & 0 (0) & 338 (6) & .055 \\
 & .25  & 0  &  & 45  & 45  & .25  &  & 69 (34) & 69 (34) & .251  &  & 45 (25) & 45 (25) & .250 \\
 &  & .2  &  & 0  & 209  & .17  &  & 33 (19) & 211 (6) & .166  &  & 0 (2) & 209 (5) & .165 \\
 &  & .4  &  & 0 & 322  & .10  &  & 26 (14) & 323 (6) & .100  &  & 0 (0) & 322 (6) & .099 \\
 & .4  & 0  &  & 259  & 259  & .35  &  & 257 (34) & 257 (34) & .355  &  & 259 (15)  & 259 (15) & .353 \\
 &  & .2  &  & 215  & 321  & .26  &  & 206 (28) & 312 (29) & .259  &  & 207 (15) & 312 (15) & .258 \\
 &  & .4  &  & 159  & 379  & .17  &  & 148 (31) & 370 (30) & .171  &  & 149 (16) & 369 (15) & .170 \\
(B-2) & .15  & 0  &  & 241  & 241  & .13  &  & 241 (31) & 241 (31) & .127  &  & 242 (13) & 242 (13) & .126 \\
 &  & .2  &  & 99  & 383  & .05 &  & 103 (37) & 391 (19) & .047  &  & 97 (20) & 386 (11) & .047 \\
 &  & .4  &  & 0  & 619  & .01  &  & 38 (19) & 622 (14) & .011  &  & 0 (0) & 620 (13) & .010 \\
 & .25  & 0  &  & 344  & 344  & .16  &  & 345 (30) & 345 (30) & .165  &  & 345 (10)  & 345 (10) & .164 \\
 &  & .2  &  & 234  & 452  & .08  &  & 237 (22) & 455 (25) & .082  &  & 236 (10) & 454 (11) & .081 \\
 &  & .4  &  & 112  & 577  & .03  &  & 111 (33) & 581 (26) & .028  &  & 111 (13) & 578 (11) & .028 \\
 & .4  & 0  &  & 457  & 457  & .18  &  & 458 (30) & 458 (30) & .184  &  & 458 (9) & 458 (9) & .183 \\
 &  & .2  &  & 344  & 564  & .10  &  & 347 (20) & 567 (30) & .100  &  & 347 (8) & 568 (12) & .100 \\
 &  & .4  &  & 237  & 671  & .05  &  & 240 (19) & 677 (38) & .047  &  & 239 (8) & 676 (13)  & .046 \\
\bottomrule 
\end{tabular}
}
\end{table}

\subsection{Convergence Rate and Efficiency}
\label{sub:Convergence-Rate-and}

The second aim of our simulation studies is to examine the relative efficiency of the nonparametric and semiparametric approaches.  For this aim, we consider only the parameter setup of (B-2) with $p = .25$ and $\phi = .20$, but simulate the training data with increasing sample sizes of $n = (250, 500, 1000, 2000, 5000, 10000)$.  For each sample size, we simulate 1000 training datasets, and for each data set, we estimate the optimal rules both nonparametrically and semiparametrically.

\begin{figure}
  \caption{Large-sample convergence properties of estimated optimal cut-off boundaries.  Horizontal lines are added to indicate the sample sizes needed to achieve $\sigma_{n}=25$.}
  \label{fig:A-simulation-study}
  \begin{centering}
    \includegraphics[width=0.9\textwidth]{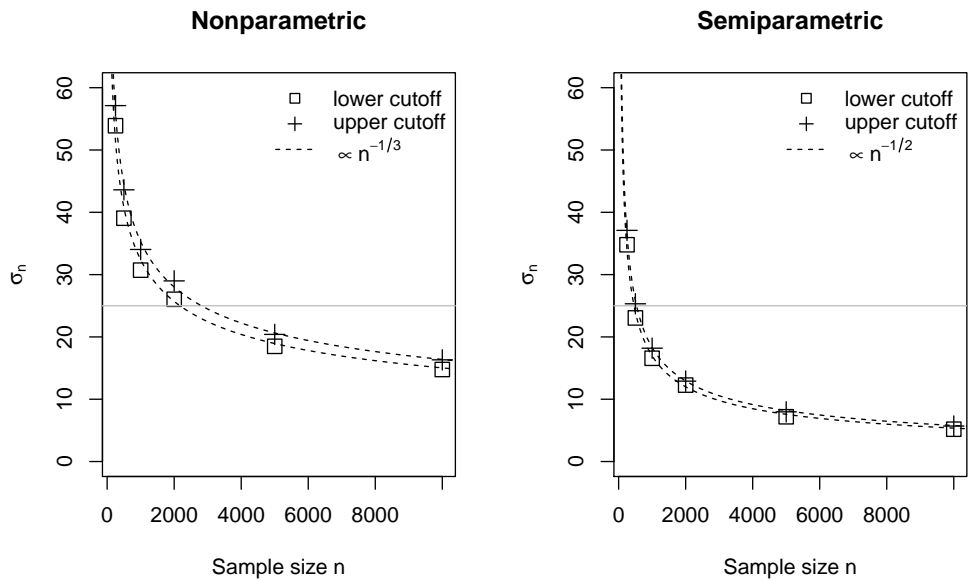}
    \par\end{centering}
\end{figure}

With a slight abuse of notation, we use $\sigma_{n}^{2}$ to denote the variances of both estimated upper and lower cut-off values. Then assuming that $\sigma_n\propto n^{-w}$ (a sufficient condition for $\sigma_{n} = O(n^{-w})$ as $n \rightarrow \infty$), we use simulations to approximate $w$ for the two estimation approaches.  Specifically, we compute $\sigma_{n}$ based on the 1000 estimated optimal cut-off values for each sample size $n$.  We then regress $\log(\sigma_{n})$ on $(-\log n)$ using a simple linear model with a slope $w$.  By least-squares estimation, we find that $w \approx 0.33$ when the optimal cut-off values are estimated nonparametrically, and $\approx 0.50$ when they are estimated semiparametrically.  The results are shown in Figure~\ref{fig:A-simulation-study}.

The simulations suggest that in this specific case, the nonparametric estimates of the optimal cut-off values converge approximately at a rate of $O(n^{-1/3})$, and the semiparametric estimates converge at a faster rate of about $O(n^{-1/2})$.  The relative efficiency between the two estimation approaches is approximately $O(n^{1/6})$ when $n$ is large.

\subsection{Simulation-Based Study Design}

The results above also suggest that a study for tripartite rule development can be designed based on simulations.  For example, suppose that the same assumptions as in Section \ref{sub:Convergence-Rate-and} are made, and we would like to design a study to determine an optimal tripartite rule such that the 95\% confidence intervals of both upper and lower cut-offs have widths of no more than $100$ CD4 (i.e., $\sigma_{n} \le 25$).  Then referring to the gray horizontal lines in Figure~\ref{fig:A-simulation-study}, a study with a sample size of about 3000 subjects is needed if the nonparametric approach is used to estimate the optimal rule, or a sample size of about 500 if the exponential tilt assumption holds and the semiparametric approach is used.

\section{Application}
\label{sec:Application}

\subsection{Data from the Miriam Hospital Immunology Clinic}
\label{sub:ICDBdata}

For illustration, we analyze data from the Miriam Immunology Clinic in Providence, RI, USA, the largest HIV clinic in the state \citep{Gillani2009}.  We recognize the essential difference between HIV-infected patients in the US and RLS.  The main reason we use a US dataset to demonstrate the development of clinical rules is because this database contains high quality CD4 and VL data that were routinely and frequently collected.

We use data from the most recent clinic visits of 597 patients who meet the following criteria: have taken ART for at least 6 months; have CD4 count, CD4\% and VL measure available at the most recent clinic visit; and have CD4 count and CD4\% available 6 months (with a window of $6\pm1$ mo) prior to that visit.  We calculate the 6-month changes in CD4 count and in CD4\%, where [6-month change] is defined as ([current] $-$ [6-mo ago])/[6-mo ago].  Total time on ART, while a potentially important predictor \citep{Kantor2009}, is not available for all patients and therefore not used in formulating risk scores.

Table \ref{tab:Summary} provides summary statistics for key clinical and immunological markers in the data.  For uniformity, viral failure is defined as having VL above 400 copies/mL (some of the VL test assays used have lower detection limits of $<400$ copies/mL). Among the 597 patients, 146 have viral failure; so the estimated prevalence of viral failure is $\widehat{p}=146/597=.25$.

\begin{table}
  \caption{Summary statistics for key variables ($n$ = 597)}
  \label{tab:Summary}
  \begin{centering}
    \begin{tabular}{llcccc}
      \toprule 
      \multicolumn{2}{c}{Marker} & mean & median & IQR & range\\
      \midrule 
      \multicolumn{2}{l}{\emph{Virological marker}} &  &  &  & \\
      & VL at most recent visit (copies/mL) & $11.8$K & 75 & (75, 400) & (12, $>500$K)\\
      \multicolumn{2}{l}{\emph{Immunological markers}} &  &  &  & \\
 &  CD4 count at most recent visit (cells/uL)  & 442 & 407 & (254, 576) & (8, $1412$)\\
 & 6-month CD4 count change (\%) & 7.3 & 18 & ($-$13, 33) & ($-$80, 736)\\
 & CD4 \% at most recent visit & 24 & 23 & (17, 30) & (.90, $59$)\\
 & 6-month CD4\% change (\%) & 9.5 & 4.7 & ($-$6.1, 16) & ($-$74, 209)\\
\bottomrule
\multicolumn{2}{l}{K: thousand; IQR: Interquartile range.} &  &  &  & \\
\end{tabular}
\par\end{centering}
\end{table}

\subsection{Risk Scores}

Two risk scores are considered for developing diagnostic rules.  The first risk score is $S_1=-\mathrm{CD4}$, or negative value of the most recent CD4 count (to be consistent with the notion that greater values of $S$ correlate with increased risk of viral failure).  The second risk score is a prediction-based composite score derived from a logistic regression of treatment failure on four immunological markers as follows,
\begin{equation*}
  S_{2}=\textrm{logit}^{-1}(\beta_{0}+\beta_{1}\textrm{[CD4]}+
  \beta_{2}\textrm{[CD4 \%]}+ \beta_{3}\textrm{[6-mo CD4 change]}+ 
  \beta_{4}\textrm{[6-mo CD4\% change]}),
\end{equation*}
where CD4 and CD4\% refer to their measures at current visit.  The MLEs (SEs) of the coefficients are $\widehat{\beta}_{0} = .89$ (.27), $\widehat{\beta}_{1} = -.0021$ (.00074), $\widehat{\beta}_{2} = -.049$ (.017), $\widehat{\beta}_{3} = -.055$ (.21), and $\widehat{\beta}_{4} = -1.40$ (.46).  A Hosmer-Lemeshow test gives a p-value of .28, indicating no evidence of lack of fit.  The distribution of $S_{2}$ has a median .21, ranges from .01 to .87, and can be interpreted as the predicted probability of treatment failure.

The risk score $S_1$ is easier to implement in clinical practice but known to have a high error rate for diagnosing viral failure.  By incorporating more clinical information, $S_2$ is potentially more accurate, but its use in clinical settings is not as straightforward as $S_1$.

\subsection{Two Simple Rules}
\label{sec:two-simple-rules}

Before calculating tripartite rules based on criteria laid out in Section \ref{sec:Optimal-Rule-Selection}, we summarize operating characteristics of two simple diagnostic rules that are similar in spirit to those commonly used in RLS when VL test has limited or no availability.

The first rule assumes that no VL testing is available (i.e., $\phi=0$) and uses CD4 $< 200$ as the hard cut-off for diagnosing treatment failure and CD4 $\ge$ 200 as non-failing, a criterion recommended by the WHO for the RLS \citep{WHO2010}.  (Another criterion recommended by the WHO for the RLS is using CD4 = 350 as the cut-off threshold.)

The second rule assumes that the limited VL testing will be used only as a confirmative test for patients with CD4~$< 200$.  This rule classifies those with CD4 count $\ge$ 200 as non-failure, and makes correct diagnoses for patients with CD4 count~$<200$.  In the Miriam Immunology Clinic data, about 15\% of patients have current CD4 count less than 200, so we consider the case that VL testing is available at 15\% of patient visits, i.e., $\phi=.15$.

The diagnostic accuracies of these two rules are summarized in Table \ref{tab:empiricalrules}.  Both rules have FNR around 0.70.  The second rule, by having 15\% of patients tested for VL, reduces the FPR to 0 and TMR from .26 to .18.  The improvement realized by having VL testing available to a small fraction of patients is evident; however, whether the second rule is optimal needs further investigations.

\begin{table}
  \caption{Diagnostic accuracies of the two simple empirical rules.  
    \label{tab:empiricalrules}}
  \centering{}
  \begin{tabular}{ccccccc}
    \toprule 
    & \multicolumn{3}{c}{Diagnosis action based on CD4} &  &  & \\
    \cmidrule{2-4} 
    $\phi$ & test positive & request VL test & test negative & 
    FPR & FNR & TMR\\
    \midrule
    0 & 0 - 200 & -- & $\ge 200$ & .10  & .70 & .26\\
    .15 & -- & 0 - 200 & $\ge 200$ & 0  & .70  & .18 \\
    \bottomrule
\end{tabular}
\end{table}

\subsection{Analysis I: CD4-Based Min-$\lambda$ Rules}

In this section, we evaluate the diagnostic performance of optimal tripartite rules based on $S_1$, using $R_{3}(\cdot)$ as the risk criterion.  To make a direct comparison to the simple rules in the last section, we assume that VL testing is available at 15\% of patient visits.  The optimal tripartite rules are developed using the nonparametric approach as described in Section \ref{sec:Optimal-Rule-Selection}.

\begin{figure}
  \caption{The optimal min-$\lambda$ rules based on $S_{1}$ and associated FPR and FNR.}
  \label{fig:The-optimal-rules-a}
  \begin{centering}
    \includegraphics*[width=0.8\textwidth]{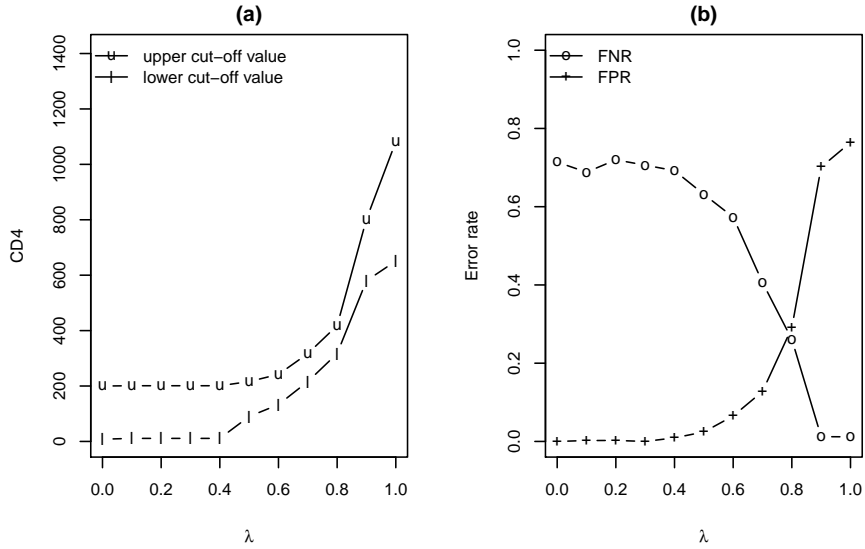}
    \par\end{centering}
\end{figure}

Figure \ref{fig:The-optimal-rules-a} shows the estimated optimal rules and associated FNR and FPR, for $\lambda$ varying from 0 to 1.  The FNR and FPR are computed using 10-fold cross validations, carried out as follows. We randomly subdivide the data into 10 subsets of about equal size; determine the FPR and FNR for each subset using the optimal rule developed using the remaining 9 subsets; and then calculate the FPR and FNR as the averages over the 10 pieces \citep[cf.][]{Hastie2001}.

As shown in Figure \ref{fig:The-optimal-rules-a}, when $\lambda$ increases (i.e., placing higher priority on avoiding false negative diagnoses), the estimated optimal rule shifts gradually toward triaging the VL tests to those with CD4 count in the middle and high range.  At $\lambda=.8$, the estimated optimal rule calls for testing patients having CD4 count between 300 and 450; in this case, both FNR and FPR are around .30.  At the extreme when $\lambda=1$, the estimated optimal rule calls for VL tests on those with CD4 $> 650$, which reduces the FNR to $\approx 0$ but increases FPR to $\approx .80$.

The left panel of Figure \ref{fig:The-optimal-rules-a} shows that when $\lambda<.4$, the estimated optimal rule is to obtain VL when $17<$ CD4 $<201$. That is, when avoidance of false positive diagnosis is prioritized, the simple rule using VL testing as a confirmative test is optimal and a reasonable choice.


\subsection{Analysis II: Comparing $S_{1}$- and $S_{2}$-Based Rules that Minimize the Weighted Risk $R_{3}$ }
\label{sec:analys-ii}

Next, we compare the diagnostic accuracy of single-maker tripartite rules based on $S_{1}$ to multiple-marker rules based on $S_{2}$ using $R_{3}(\cdot)$ as the risk criterion.  We consider three values of $\lambda=(.25,.50,.75)$ and three constraints on VL test availability $\phi=(0,.15,.30)$.  Nonparametric estimates of the optimal rules, along with FPR, FNR and TMR obtained from cross-validations, are given in Table \ref{tab:Comparing.rules}.  Standard errors for all table entries are computed using the bootstrap method with 500 re-samples.

In summary, the optimal rules based on $S_{2}$ have a slightly better diagnostic performance than the optimal rules on $S_{1}$.  However, the magnitude of improvement by incorporating more non-VL markers is small, relative to the improvement that can be achieved by the selective use of VL testing on more patients. In Section \ref{sub:Analysis-ROC}, the diagnostic accuracies of the rules based on the two risk scores will be further compared using AUCs.

\begin{table}
  \caption{Comparison of the $S_{1}$- and $S_{2}$-based tripartite rules.  The optimal cut-off points based on $S_{1}$ are transformed back to the original scale of CD4 count.  The numbers in parentheses are standard errors.} \label{tab:Comparing.rules}
  \centering{}
  \begin{tabular}{ccccccccccc}
    \toprule 
    &  &  &  & \multicolumn{2}{c}{cut-off points} &  &  &  &  & \\
    \cmidrule{5-6} 
    & $\lambda$ & $\phi$ &  & lower & upper &  & FNR  & FPR  & $R_{3}$ & TMR \\
    \midrule
    $S_{1}$ & .25  & .00  &  & 17 (17) & 17 (17) &  & .98 (.02)  & .00 (.00)  & .06 (.01) & .24 (.02) \\
    &  & .15  &  & 17 (9) & 201 (18) &  & .72 (.04) & .00 (.00) & .04 (.00) & .18 (.01)\\
    &  & .30  &  & 17 (13) & 284 (13) &  & .46 (.04) & .00 (.00) & .03 (.00) & .12 (.01)\\
    & .50  & .00  &  & 120 (64) & 120 (64) &  & .93 (.12) & .03 (.04) & .12 (.01) & .26 (.02)\\
    &  & .15  &  & 90 (16) & 216 (39) &  & .63 (.06) & .02 (.01) & .08 (.01) & .17 (.01)\\
    &  & .30  &  & 17 (15) & 284 (32) &  & .45 (.04) & .01 (.01) & .05 (.01) & .12 (.01)\\
    & .75  & .00  &  & 302 (45) & 302 (45) &  & .43 (.06) & .26 (.06) & .13 (.01) & .30 (.04)\\
    &  & .15  &  & 216 (51) & 317 (50) &  & .40 (.07) & .13 (.06) & .10 (.01) & .20 (.04)\\
    &  & .30  &  & 226 (67) & 417 (81) &  & .30 (.07) & .14 (.06) & .08 (.01) & .18 (.04) \\
    \midrule
    $S_{2}$ & .25  & .00  &  & .64 (.04) & .64 (.04) &  & .91 (.04) & .01 (.00) & .06 (.00)  & .23 (.01)\\
    &  & .15  &  & .39 (.01) & .75 (.04) &  & .62 (.04) & .00 (.00)  & .04 (.00) & .16 (.01)\\
    &  & .30  &  & .29 (.01) & .71 (.04) &  & .42 (.04) & .00 (.00) & .03 (.00) & .11 (.01)\\
    & .50  & .00  &  & .53 (.07) & .53 (.07) &  & .79 (.08) & .04 (.02) & .11 (.01) & .23 (.01)\\
    &  & .15  &  & .37 (.01) & .66 (.06) &  & .60 (.05) & .01 (.01) & .07 (.01) & .16 (.01)\\
    &  & .30  &  & .28 (.01) & .67 (.06) &  & .43 (.04) & .01 (.01) & .05 (.01) & .11 (.01) \\
    & .75  & .00  &  & .26 (.04) & .26 (.04) &  & .35 (.08) & .30 (.08) & .12 (.01) & .31 (.04)\\
    &  & .15  &  & .26 (.04) & .34 (.05) &  & .32 (.08) & .21 (.07) & .09 (.01) & .24 (.03)\\
    &  & .30  &  & .19 (.03) & .34 (.08) &  & .25 (.06) & .13 (.05) & .07 (.01) & .16 (.03)\\
    \bottomrule
\end{tabular}
\end{table}

\subsection{Analysis III: Optimal Rules under Exponential Tilt Assumption}
\label{sec:analysis3.exp.tilt}

In this section, we develop the optimal tripartite rules under the exponential tilt assumption.  We consider the following two risk scores, $S_{1}$ and $S_{1}^{*}=-\log_{10} (\text{CD4)}$, for rule development.  The reason for using $S_{1}^{*}$ is that it avoids the issue of having the empirical adjustment when one cut-off value is beyond the support of the risk score (as we encountered in our simulation studies). The risk score $S_1^*$ may also be more suitable for the exponential tilt model.

\begin{figure}
  \caption{Empirical and semiparametric estimates of the cumulative densities of CD4 counts and $\log_{10}$(CD4).}
  \label{fig:Cumulative-densities}
  \begin{centering}
    \includegraphics*[width=.9\textwidth]{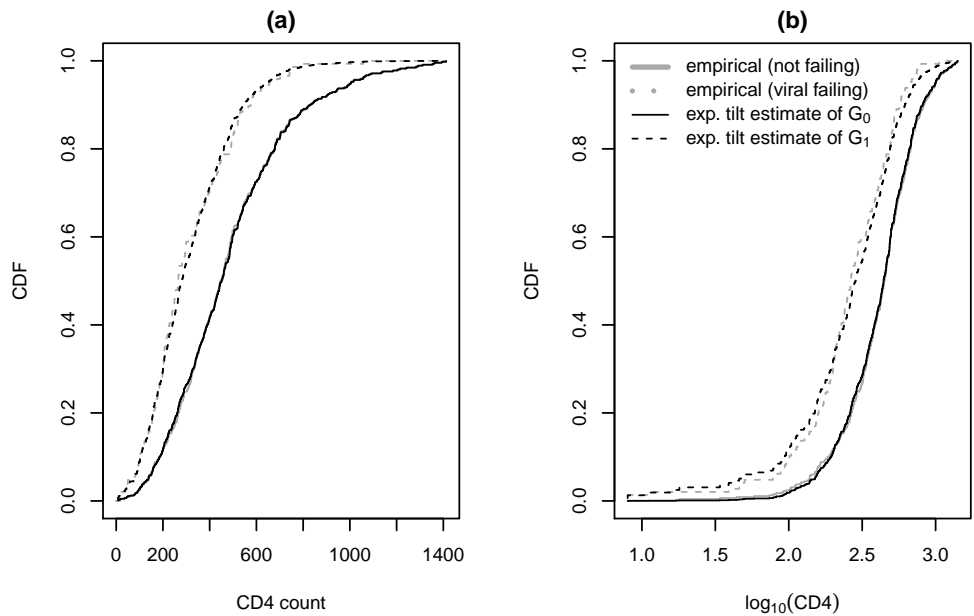}
    \par\end{centering}
\end{figure}

We first examine the suitability of the exponential tilt model for $S_{1}$ and $S_{1}^{*}$ by plotting the semiparametric estimates of $G_{0}$ and $G_{1}$ against their empirical estimates.  The results are shown in Figure \ref{fig:Cumulative-densities}, where the semiparametric estimates of $G_{0}$ and $G_{1}$ are obtained using the results in Appendix \ref{sub:A.-Semi-parametric-estimate}. Figure~\ref{fig:Cumulative-densities} suggests that the exponential tilt assumption is reasonable for both $S_{1}$ and $S_{1}^{*}$ although the goodness of fit for $S_1$ is slightly better.  (One also can use Q-Q plots, not shown, to examine the model goodness of fit.)

The estimated optimal rules  using TMR as the risk criterion are given in Table \ref{tab:Exponential-tile-optimal-rules}.  The intervals for triaging VL assays are centered at CD4 = 77 and 109 for the optimal rules based on $S_{1}$ and $S_{1}^{*}$, respectively.  Overall, the diagnostic accuracies of the two sets of estimated optimal rules are comparable, and their estimated cut-off values differ only slightly relative to their standard errors.  The optimal rules in Table~\ref{tab:Exponential-tile-optimal-rules} also are comparable to the optimal rules that are developed nonparametrically (see Table \ref{tab:Comparing.rules} with $\lambda=.50$), but in general have much smaller standard errors.

\begin{table}
  \caption{The optimal min-TMR rules under the exponential tilt assumption.  Cut-off points were transformed back to the original scale of CD4 count. The numbers in parentheses are standard errors calculated using the bootstrap method. }
  \label{tab:Exponential-tile-optimal-rules} 
  \centering
  \begin{tabular}{ccccccc}
    \toprule 
    &  & \multicolumn{2}{c}{cut-off points} &  &  & \\
    \cmidrule{3-4} 
    & $\phi$ & lower & upper & FNR & FPR & TMR \\
    \midrule 
    $S_{1}$ & 0 & 77 (46) & 77 (46) & .94 (.06) & .01 (.02) & .24 (.02)\\
    & .15 & 0 (29) & 199 (10) & 69 (.05) & .00 (.01) & .17 (.02)\\
    & .30 & 0 (10) & 278 (11) & .47 (.04) & .00 (.00) & .11 (.01)\\
    \midrule
    $S_{1}^{*}$ & 0 & 109 (23) & 109 (23) & .89 (.05) & .03 (.01)  & .24 (.02)\\
    & .15 & 57 (22) & 206 (10) & .67 (.04) & .01 (.01) & .18 (.02)\\
    & .30 & 41 (17) & 287 (12) & .46 (.04) & .01 (.00) & .12 (.01)\\
    \bottomrule
\end{tabular}
\end{table}

\subsection{Analysis IV: ROC Analyses for Tripartite Rules}
\label{sub:Analysis-ROC}

\begin{figure}
  \caption{ROC curves for diagnostic rules using $S_{1}$ and $S_{2}$ (subplots (a) and (b)); the resulting AUC curves as functions of $\phi$ (subplot (c)); and the difference of the two AUC curves ($S_{2}$ ``minus'' $S_{1}$, subplot (d)).  The point-wise 95\% CI of the difference in AUC is calculated using the bootstrap method with 500 re-samples.} \label{fig:ROC-curves-of}
  \begin{centering}
    \includegraphics*[width=.95\textwidth]{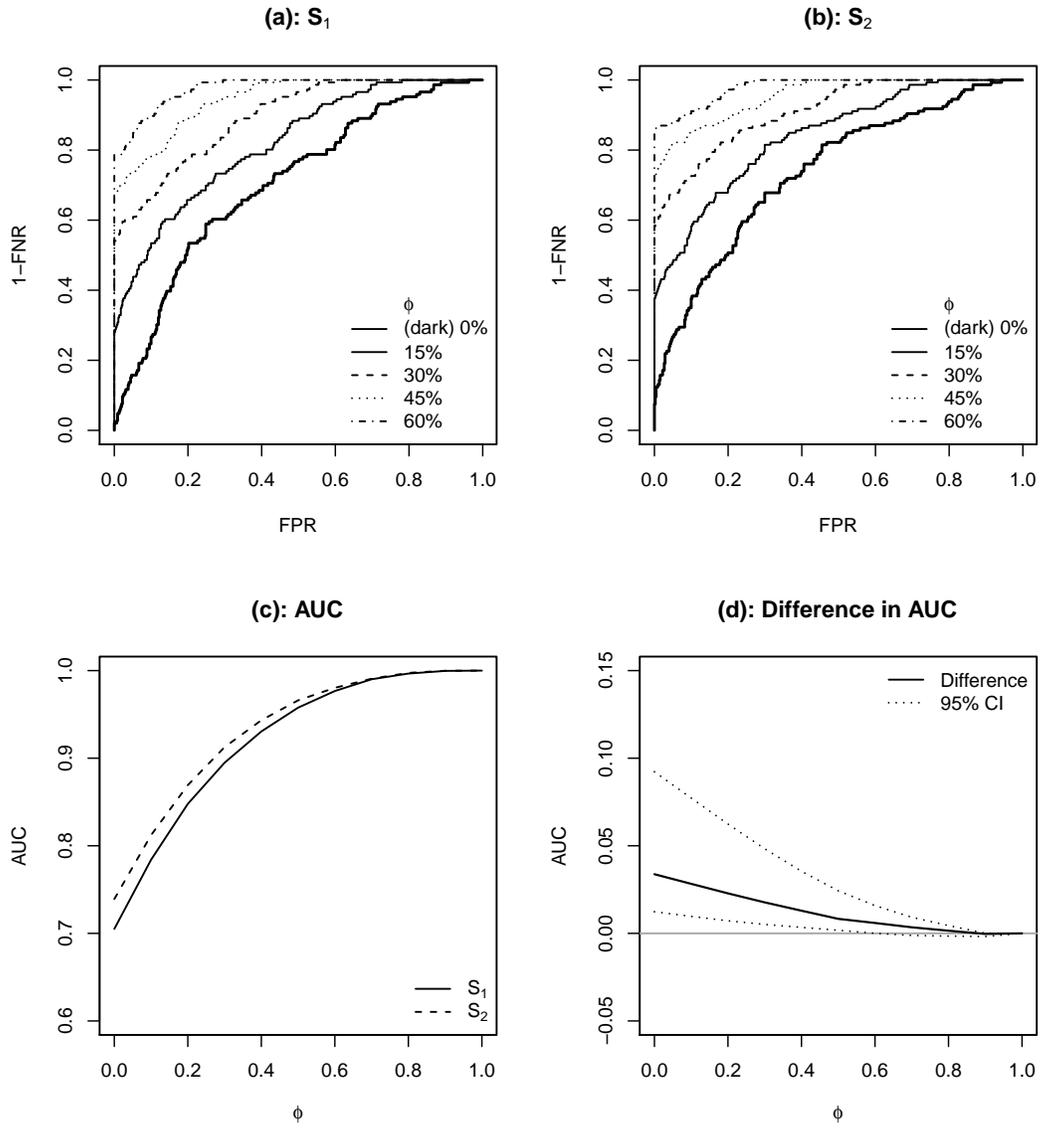}
    \par\end{centering}
\end{figure}

Figure \ref{fig:ROC-curves-of} shows nonparametric estimates of ROC curves for tripartite rules based on $S_{1}$ and $S_{2}$, when the VL tests are available at $\phi=0$, 15, 30, 45, and 60\% of patient visits.  The ROC curves in the subplot (b) are slightly better than those in the subplot (a), which suggests that improvement in diagnostic capacity can be achieved using the composite score $S_{2}$.  See also the AUC curves and their difference in the subplots (c) and (d).  Consistent with our findings in Analysis II (Section \ref{sec:analys-ii}), the difference between the two AUC curves, although statistically significant for $\phi<.6$, is marginal.

Relative to not having VL tests available, the AUCs for tripartite rules based on both risk scores are substantially improved as VL testing is made available for some of clinical visits.  For example, as shown in the subplot (c), when we increase the VL test availability from 0 to 20\%, the absolute improvement in AUC is about 15\%; and increasing availability to 40\% improves AUC by more than 20\%.  In particular, the \emph{relative} improvement by making VL testing accessible to some HIV patients is more pronounced when the VL testing availability is low.


\section{Discussion}
\label{sec:Discussion}

This paper is motivated by recent evidence that the CD4-based WHO guidelines for monitoring HIV treatment in RLS can lead to high treatment failure misclassification rates, and by the fact that VL tests are becoming available to programs and patients in RLS, typically on a limited basis.  To make optimal use of VL tests, we propose a tripartite diagnostic rule based on a risk score that subdivides patients into a high-risk group (classified as treatment failure), a low-risk group (as viral suppressed), and an intermediate-risk group to whom the limited VL tests are assigned, where the size of the third group is constrained by the availability of VL tests.  Nonparametric and semiparametric methods are proposed for determining an optimal rule to minimize a given risk criterion.  ROC analysis procedure for characterizing the diagnostic performance of tripartite rules and its associated asymptotic properties are developed.

Our proposed method is demonstrated by analyzing data from the Miriam Hospital Immunology Clinic in Providence, RI.  We show that with selective and targeted use of VL tests, the rate of misdiagnosis can be substantially reduced even when VL testing is available at a small portion of patient visits (e.g., $\phi=15\%$).  Our analysis also suggests that when avoidance of false positive diagnoses is prioritized, using VL testing strictly to confirm viral failure for those deemed to be at high risk is a reasonable choice.  This finding applies only to patients at the Miriam Hospital Immunology Clinic; its external validity remains to be tested.

Our methods assume that the functional form of risk score $S(\mathbf{X})$ is known, but may be relaxed by unknown parameters.  When the function form of $S(\mathbf{X})$ is unknown, methods of machine/statistical learning, such as boosting \citep{Freund:1999}, targeted/super learner \citep{Sinisi2007, Van-Der-Laan2011}, classification tree learning \citep{Breiman:1984}, neural networks \citep{Hagan:1996, Sarle:1994}, and prediction-based classification methods \citep{Foulkes2002}, can be implemented. We refer the readers to \citet{Hastie2001} and \citet{Kotsiantis:2007} for a more comprehensive treatment on the topic.

We assume that there is no measurement error in VL, i.e.\ that VL test is the gold standard for determining the amount of circulating virus.  In developed countries, repeated VL tests and HIV genotyping are usually required to confirm treatment failure and existence of drug resistance once the VL becomes detectable.  HIV-infected patients in RLS however do not have such luxury, and a single VL test result (if done) is probably the most direct measure of viral failure, and is used for clinical decision making.  So although the assumption is not ideal, it is reasonable in this context because the measured VL is the best available basis for decision making in RLS. Future work will address the issue of measurement error in VL and its effect on misclassification rates.

CD4 counts are known to be highly variable due to measurement error, diurnal variations, and other factors.  The measurement error of CD4 count may be part of the reason for high misclassification rates of the WHO guidelines.  The impact of measurement error in biomarkers on predicting binary outcomes has been studied by \citet{Carroll:1984,Carroll:2006}, \citet{Buzas:2003}, and \citet{Fuller:2009} among others.  Generally speaking, large measurement errors of a biomarker are associated with a greater attenuation of its capability of predicting outcomes.  One way to reduce the impact of measurement errors is through repeated measurements.  Given the fact that point-of-care CD4 technologies are being developed, it may be possible in practice in the future to quantify and reduce the impact of CD4 measurement error by multiple testing at a single visit.  On the other hand, with additional information such as prior history of CD counts, it may also be possible to evaluate the magnitude of measurement error by constructing appropriate measurement error models (which typically rely on certain subjective assumptions) and applying methods such as regression calibration \citep{Carroll:1990, Rosner:1990} and simulation-extrapolation \citep{Stefanski:1995}.  Improving diagnostic accuracy by reducing the impact of CD4 measurement error is an area worthwhile further investigation.

A final limitation of this paper, as it applies to developing rules for RLS, is that a US data set is used to demonstrate our proposed methods.  Our ongoing work is focused on developing and calibrating rules based on data from sub-Saharan Africa and other RLS.

~

\bibliographystyle{asa}
\bibliography{refs}

\appendix
\makeatletter 
\renewcommand{\@seccntformat}[1]{APPENDIX~{\csname
    the#1\endcsname}.\hspace*{1em}} \makeatother
\section{}

\subsection{Semiparametric estimates of $G_0$, $G_1$, and $G$ under the exponential tilt assumption}
\label{sub:A.-Semi-parametric-estimate}

Suppose that we want to estimate the mixture distribution $G$ using an i.i.d\ sample of $\{(S_{i},Z_{i}): i=1,\dots,n\}$.  In the spirit of nonparametric likelihood estimation, we consider only the distributions with jumps at $\{S_{i}\}$.  Thus the (profile) likelihood for $ G_0$ can be written as \citep[see][]{Qin1999}
\begin{align*}
  \mathbf{L}(G_{0}) & \propto \prod_{\{Z_{i}=0\}}g_{0}(S_{i})\prod_{\{Z_{i}=1\}}
  \exp(\widehat{\beta}_{0}^{*}+\widehat{\beta}_{1}S_i)g_{0}(S_{i}) \\
  &= \{\prod_{i=1}^{n}\theta_{i}\}\prod_{\{Z_{i}=1\}}\exp(\widehat{\beta}_{0}^{*}
  +\widehat{\beta}_{1}S_{i}),  
\end{align*}
where $\widehat{\beta}_{0}^{*} = \widehat{\beta}_{0}- \mathrm{logit}(\widehat{p})$, $\widehat{\beta}_{0}$ and $\widehat{\beta}_{1}$ are the MLEs from the logistic regression~\eqref{eq:logistic.model}, and $\theta_{i} = g_{0}(S_{i})$ denotes the mass at the observed $S_{i}$ with $\sum_{i}\theta_{i}=1$.  Here we proceed as if we have $n$ distinct values in $\{S_{i}\}$, which does not affect the following results.  Applying the Lagrange multiplier, one can show that the likelihood is maximized at 
$$\widehat{\theta}_{i} = \frac{1}{n}\left[1 + \nu\{\exp(\widehat{\beta}_{0}^{*} + \widehat{\beta}_{1}S_{i})-1\}\right]^{-1},$$ 
where $\nu$ is the Lagrange multiplier solving
\begin{equation*}
  \sum_{i=1}^{n}\frac{\exp(\widehat{\beta}_{0}^{*}+
    \widehat{\beta}_{1}S_{i})-1}{1+\nu\{\exp(\widehat{\beta}_{0}^{*}+
    \widehat{\beta}_{1}S_{i})-1\}}=0.
\end{equation*}
We then estimate $G_{0}$, $G_1$, and $ G$ semiparametrically by
\begin{align*}
 & \widetilde{G}_{0}(s)=\sum_{i=1}^{n}\widehat{\theta}_{i}\mathbf{1}\{S_{i}<s\}, \\ \nonumber 
 & \widetilde{G}_{1}(s)=\sum_{i=1}^{n}e^{(\widehat{\beta}_{0}^{*}+
    \widehat{\beta}_{1}S_{i})}\widehat{\theta}\mathbf{1}\{S_{i}<s\}, \\ \nonumber
 & \widetilde{G}(s)=(1-\widehat{p})\widetilde{G}_{0}+\widetilde{p}\widetilde{G}_{1}.
\end{align*}
Because the exponential tilt assumption places no constraint on the marginal distribution $G$, it can be verified that the semiparametric estimate $\widetilde{G}(s)$ is equal to the empirical estimate $\widehat{G}(s)$.  

\subsection{Properties of AUC$_{\phi}$ for tripartite rules}
\label{sub:Properties-of-AUC}

\noindent \textbf{Property A.2.1}: {\it Let $S\sim G_1$, $S'\sim G_0$, and $S$ and $S'$ be independent.  Then,
\begin{equation}
  \label{eq:AUC.result.1a}
  \mathrm{AUC}_{\phi}=\Pr\{S>H_{\phi}(S')\}+\frac{1}{2}\Pr\{S=H_{\phi}(S')\},
\end{equation}
and 
\begin{equation}
  \label{eq:AUC.result.1b}
  \mathrm{AUC}_{\phi}=\frac{1}{2}\left(1+\mathrm{E}_{G_1}[ G_0(S)]-
    \mathrm{E}_{ G_0}[G_1\{H_{\phi}(S')\}]\right).
\end{equation}}

\noindent Proof: We have 
\begin{align*}
  \textrm{AUC}_{\phi} & =
  \int_{\infty}^{-\infty}\{1-G_1(H_{\phi}(u))\}
  \mathrm{d}(1- G_0(u)) \\
  & =\int_{-\infty}^{\infty}\Pr(S >H_{\phi}(u)) \mathrm{d} G_0(u) \\
  & \Rightarrow  \eqref{eq:AUC.result.1a},
\end{align*}
where $\frac{1}{2}\Pr(X=H_{\phi}(X'))$ is added for ties and the term vanishes for continuous $S$.  Further,
\begin{align*}
  \textrm{AUC}_{\phi} & =1-\int_{-\infty}^{\infty}G_1(H_{\phi}(u))
  \mathrm{d} G_0(u) \\
  & =\int_{-\infty}^{\infty} G_0(u)\mathrm{d}
  G_1(H_{\phi}(u)) \\
  & \Rightarrow \eqref{eq:AUC.result.1b}.  
\end{align*}
$\Box$

\noindent \textbf{Property A.2.2}: {\it If $S$ is stochastically greater than $S'$, then AUC$_{\phi}$ is bounded by
\begin{equation*}
  \frac{1}{2}+\phi-\frac{\phi^{2}}{2}\le\textrm{AUC}_{\phi}\le1.
\end{equation*}
The lower bound is achieved when $G_1= G_0$, and the upper bound when $\phi=1$. }  

\noindent Proof: We prove the results for the case when $S$ is continuous such that there exist $l$ and $u$ with $ G(u)- G(l)=\phi$.  Manipulating this constraint slightly, we have $1- G(l)=1- G(u) + \phi\Rightarrow1-\{pG_1(H_{\phi}(u))+(1-p) G_0(H_{\phi}(u))\} = 1-\{pG_1(u)+(1-p) G_0(u)\}+\phi$.  The condition that $S$ is stochastically greater than $S'$ implies that
\begin{align*}
  1-G_1(H_{\phi}(u)) &\ge1-\{pG_1(u)+(1-p) G_0(u)\}+ \phi \\
  &\ge1- G_0(u)+\phi.
\end{align*}
Therefore, 
\begin{align*}
  \textrm{AUC}_{\phi} & =\int_{u=\infty} ^{-\infty}\{1-G_1(H_{\phi}(u))\}
  \mathrm{d}(1- G_0(u)) \\
  & \ge\int_{u=\infty}^{-\infty}[\{1- G_0(u)+\phi\}
  \wedge1]\mathrm{d}(1- G_0(u))\\
  & =\int_{u=\infty}^{ G_0^{-1}(\phi)}\{1- G_0(u)+\phi\}\mathrm{d}
  (1- G_0(u))+\int_{u= G_0^{-1}(\phi)}^{-\infty}\mathrm{d}
  (1- G_0(u))\\
  &=\frac{1}{2}+\phi-\frac{\phi^{2}}{2}.
\end{align*}
All equalities hold when $G_0=G_1$.  $\Box$

\subsection{Asymptotic properties of estimated ROC curve and AUC}
\label{sub:Asymptotic-properties}

The nonparametric estimate $\widehat{C}_{\phi}$ given by \eqref{eq:ROC.curve.est} has the following properties:

\noindent \textbf{Property A.3.1}: {\it The nonparametric estimate $\widehat{C}_{\phi}$ is uniformly consistent.}

\noindent Proof: Let us write
\begin{align*}
  \sup_t|\widehat{C}_{\phi}(t)-C_{\phi}(t)|= & \sup_{t}|\widehat{G}_{1}
 \circ\widehat{H}_{\phi}\circ\widehat{G}_{0}^{-1}(t)-G_1
 \circ\widehat{H}_{\phi}\circ\widehat{G}_{0}^{-1}(t)|\\
 & +\sup_{t}|G_1\circ\widehat{H}_{\phi}\circ\widehat{G}_{0}^{-1}(t)-
 G_1\circ H_{\phi}\circ\widehat{G}_{0}^{-1}(t)|\\
 & +\sup_{t}|G_1
 \circ H_{\phi}\circ\widehat{G}_{0}^{-1}(t)-G_1\circ H_{\phi}
 \circ G_0^{-1}(t)|.
\end{align*}
Then, it  can be shown  that the first  term converges to  zero almost surely  by the  Glivenko-Cantelli Theorem,  and the  second and  third terms converge to zero almost surely by the Law of Large Numbers.  See \citet{Hsieh1996}.  $\Box$

\noindent \textbf{Property A.3.2}: {\it Suppose that the densities $g_{0}$, $g_{1}$ and $g$ are continuous and bounded, and $\sum Z_{i}/n\rightarrow p$ as $n\rightarrow\infty$.  Then, the following approximation holds asymptotically as $n\rightarrow\infty$,
\begin{eqnarray*}
  n^{\frac{1}{2}}\{\widehat{C}_{\phi}(v)-C_{\phi}(v)\} & = & 
  \frac{1}{\sqrt{p}}B_{1}\circ C_{\phi}(v)+\frac{g_{1}\circ 
    H\circ G_{0}^{-1}(1-v)\sqrt{1-\rho}}{g\circ 
      G^{-1}\{G\circ G_{0}^{-1}(1-v)-\phi\}}B\{G\circ G_{0}^{-1}
  (1-v)-\phi\}\\
  &  & +\frac{1}{\sqrt{1-p}}\frac{g_{1}\circ H\circ 
    G_{0}^{-1}(1-v)}{g_{0}\circ G_{0}^{-1}(1-v)}\frac{g\circ 
    G_{0}^{-1}(1-v)}{g\circ G^{-1}\{G\circ G_{0}^{-1}(1-v)-
    \phi\}}B_{2}(1-v),
\end{eqnarray*}
where $B_{1}(v)$ and $B_{2}(v)$ are independent Brownian bridges, $B(v)=B_{1}(v)/\sqrt{p} +B_{2}(v)/\sqrt{1-p}$, and $\rho=\{G\circ G_{0}^{-1}(v)-\phi\}\{1-G\circ G_{0}^{-1}(v)\}/ [G\circ G_{0}^{-1}(v)\{1-G\circ G_{0}^{-1}(v)+\phi\}]$. }

The strategy for proving A.3.2 is similar to \citet{Hsieh1996}.   

\noindent \textbf{Property A.3.3}: {\it The nonparametric estimate $\widehat{\mathrm{AUC}}_{\phi}$ given by \eqref{eq:auc.nonpar.est} has the property that, as $n\rightarrow\infty$,
\begin{equation}
  \label{eq:auc.asymp.2}
  n^{1/2}\sigma^{-1/2}(\widehat{\mathrm{AUC}}_{\phi}-
  \mathrm{AUC}_{\phi})\overset{d}{\longrightarrow}\mathcal{N}(0,1),
\end{equation}
where $\sigma^2 =\mathrm{Var}_{G_1}\{G_0\circ H_{\phi}(S)\}p^{-1}+ \mathrm{Var}_{ G_0}\{G_1\circ H_{\phi}(S)\}(1-p)^{-1}.$}

\noindent Proof: We prove \eqref{eq:auc.asymp.2} using the properties of U-statistics \citep{Lee1990}.  Applying the H\'{a}jek projection principle on \eqref{eq:auc.nonpar.est} \citep{vaart1998}, we express $\widehat{\mathrm{AUC}}_{\phi}$ as
\begin{equation*}
  \label{eq:auc.asymp.1-2}
  \widehat{\mathrm{AUC}}_{\phi}=\textrm{AUC}_{\phi}+\widetilde{A}_{n}+o(1/n),
\end{equation*}
where
\begin{align*}
\widetilde{A}_{n} =& \frac{1}{\sum_{i}Z_{i}}\sum_{i}Z_{i}[\widehat{G}_{0}
\circ\widehat{H}_{\phi}(S_{i})-\mathrm{E}_{\widehat{G}_{1}}\{\widehat{G}_{0}
\circ\widehat{H}_{\phi}(S_{i})\}] \\
& + \frac{1}{\sum_{i}(1-Z_{i})}
\sum_{i}(1-Z_{i})[\widehat{G}_{1}\circ\widehat{H}_{\phi}(S_{i})-
\mathrm{E}_{\widehat{G}_{0}}\{\widehat{G}_{1}\circ\widehat{H}_{\phi}(S_{i})\}]  
\end{align*}
is a U-statistic.  Then conditional on $\sum_{i}Z_{i}$,  an ancillary statistic for the AUC, \eqref{eq:auc.asymp.2} is an immediate result of applying Slutsky's lemma and the Central Limit theorem.  $\Box$
\end{document}